\documentclass[aps,showpacs,showkeys,nofootinbib,twocolumn]{revtex4}
\usepackage{epsfig}
\usepackage{amsmath} 
\usepackage{pstricks} 

\def\Journal#1#2#3#4{{#1} {\bf #2}, #3 (#4)}

\def\NPB{{\em Nucl.~Phys.} B}
\def\NPA{{\em Nucl.~Phys.} A}
\def\PLB{{\em Phys.~Lett.}  B}
\def\PRL{\em Phys.~Rev.~Lett.~}
\def\PRD{{\em Phys.~Rev.} D}
\def\PRC{{\em Phys.~Rev.} C}

\newcommand{\mT}[1]{m_{\perp_{#1}}}
\newcommand{\pT}[1]{p_{\perp_{#1}}}
\newcommand{\kT}[1]{k_{\perp_{#1}}}

\newcommand{\pTave}[1]{\left<\pT{#1}\right>}

\newcommand{\kTave}[1]{\left<\kT{#1}\right>}

\newcommand{\Pythia}{\textsc{Pythia}}
\newcommand{\Fritiof}{\textsc{Fritiof}}

\newcommand{\UNIT}[1]{\mbox{$\,{\rm #1}$}}

\newcommand{\GeV}{\UNIT{GeV}}

\newcommand{\AGeV}{\UNIT{AGeV}}

\newcommand{\fm}{\UNIT{fm}}
\newcommand{\mb}{\UNIT{mb}}
\newcommand{\fmc}{\UNIT{fm/c}}

\newcommand{\REM}[1]{}
\newcommand{\CARSTEN}[1]{}

\newcommand{\CAPTION}[1]{\caption{#1}}

\begin{document}

\title{
Quenching of High $p_\perp$ Hadron Spectra
by Hadronic Interactions\\ in Heavy Ion Collisions at RHIC}

\author{K.~Gallmeister, C.~Greiner and Z.~Xu}

\affiliation{Institut f\"ur Theoretische Physik, %
  Universit\"at Giessen, %
  Heinrich--Buff--Ring 16, %
  D--35392 Giessen, %
  Germany}

\begin{abstract}
  Typically the materialization of high energetic
  transverse partons to hadronic jets is assumed to occur
  outside the reaction zone in a relativistic heavy ion collision.
  In contrast, a quantum mechanical estimate
  yields a time on the order of only a few fm/c for building up the
  hadronic wavefunction
  for jets with typical transverse momenta
  of $p_\perp \leq 10\GeV$ as accessible at RHIC facilities.
  The role of possible elastic or inelastic collisions of
  these high $p_\perp$ particles
  with the bulk of hadrons inside the fireball is addressed by means
  of an opacity expansion in the number of collisions.
  This analysis shows
  that the hadronic final state interactions
  can in principle account for the modification
  of the (moderate) high $p_\perp$ spectrum observed for central
  collisions at RHIC. 
\end{abstract}



\maketitle


\section{Motivation and estimates}
\label{Intro}

One of the major goals of the ongoing experiments at the Relativistic
Heavy Ion Collider (RHIC) in Brookhaven National Laboratory is to find
and probe for a temporarily occurring macroscopical state of
deconfined quark--gluon matter.
So far, especially hadronic abundancies have been experimentally
thoroughly studied.
Total hadron multiplicity measurements, however, do reflect mainly the
chemical equilibration and possible hadronization processes of the
late stages of the system, and do not deliver direct information on the
early stage with maximum energy density and temperature, where a
deconfined and highly excited state is generally expected to be formed.
Still, one of the very interesting first results of the RHIC experiments
is, that one has established a significant suppression of moderately
high $p_\perp$ hadrons produced in central $A+A$ collisions compared to
rescaled peripheral collisions or rescaled (and extrapolated) $p+p$
collisions:
The results for the lower center--of--mass energy of $130\AGeV$
\cite{PHENIX,STAR} and also the preliminary data for $200\AGeV$
\cite{QM2002} show a suppression factor $R(p_\perp)$ of about
$ \approx 1/5$ for
pions with $p_\perp\approx 5\GeV$, and hence state a clear hint
for (probably various) nuclear medium effects at work.

The most popular explanation for this phenomenon
is the onset of the occurrence of so
called `jet quenching', anticipated in \cite{Wang}. 
The idea is, that a high energetic parton moving through a dense
coloured (and deconfined) medium will loose considerable energy due to
collisions and induced gluon radiation, and its final fragmentation
will give rise to particles with considerable lower energies
\cite{Wang1}.
Hence, measurements of jets seem to offer a direct access to probe the
early stage when the deconfined matter is very dense. 
As a present theoretical consensus, the dominant mechanism for energy
loss is thought to be the nonabelian radiation of gluons based on pQCD
by the energetic partons \cite{BDMS,Baier}.  
As a result, the momenta of the jet partons are attenuated before
hadronizing. 
One should internalize the word of caution, that this appealing
picture of applying perturbative QCD calculations is strictly valid
only for such high energetic jets with $E>10\dots20\GeV$ \cite{Baier}.
And indeed, it has been found that the suppression pattern of the hard
jets (with $E>10\GeV$) can show some significant dependence on the
initial properties of the deconfined state \cite{Kai}.

A recent calculation based on the rather involved GLV formalism for a
finite number of coloured collisions  \cite{Levai} does indeed provide
a good agreement with the data. 
Still, though, a phenomenological opacity parameter has to be adjusted
\cite{Levai2}. 
In this study already various other effects (the Cronin effect as well
as a slight modification of the gluon distribution due to shadowing)
have been phenomenologically incorporated. 
These effects do partly counteract and compete with the pure parton
energy loss and thus do make a detailed analysis already more delicate
\cite{JMS02,VG02}. 
Contrary, the proposal was made that the observed spectra for central
collisions show a significant $\mT{}$ scaling, being a manifestation
of a direct remnant of the initial gluons which were liberated from
gluon saturated nuclear distribution functions \cite{SKMV01}.
Here any possible later interactions of the jets with the surrounding
medium which might or should alter the distributions are completly
discarded. 

It has also been raised very recently, that in an ideal strong
quenching picture the $p_\perp$ spectrum of the high momentum hadrons is
simply given by early `surface emission' of the outer regions, where roughly
half of the there produced jets can escape into the vacuum
without passing through any
medium \cite{Muller}. 
It has to be verified,
whether such an idealised scenario of complete absorption
of energetic partons in matter can actually be achieved.
The proposal bears resemblance to a similar
geometrical picture for the production of another hard hadronic probe,
the $J/\Psi $,
being invoked much earlier
for a possible explanation of the so called `anomalous'
suppression pattern
observed in $Pb+Pb$ collisions at CERN--SPS \cite{blaizot}.

Typically, it is assumed in all the above refered descriptions
that the partons do exit the collision region as pointlike particles
before finally fragmenting into a `jet' of hadrons.
We now will point out that this prejudice should be of
major concern, being invalid for present settings at RHIC.
The potential magnitude of the hadronization time (or, to be more
precise, the time to build up the hadronic wavefunction) is based on a
relativistic and simple quantum mechanical estimate \cite{DKMT} for
light quark hadronic states
\begin{subequations}
  \begin{align}
    t_q^{\rm hadr} &\approx E \, R^2\label{eq:1a}\\
\intertext{or heavy quark meson states}
    t_Q^{\rm hadr} &\approx \frac{E}{m_Q} \, R\quad.\label{eq:1b}
  \end{align}
\end{subequations}
Here $E$ denotes the energy of the leading `parton' or jet,
$R$ states the transversal
size of the to be formed hadrons and $m_Q$ is the mass of a heavy quark.
Taking (\ref{eq:1a}) with the average radius of the pion 
$R_\pi\approx 0.5\fm$ 
or (\ref{eq:1b}) substituting $m_Q $ by $m_\rho $ for a $\rho$ meson
and taking $R_\rho \approx 0.8\fm$, one has for the formation time
a simple understanding given as
\begin{equation}
\label{formationtime}
  t_F \, \approx \, 1 \dots 1.2 \ (E/\GeV) \cdot \fmc\quad.
\end{equation}
Hence, for leading hadrons with
moderately high $p_\perp \le 10\GeV$
original point--like jet--partons have established already a complete
nonperturbative, transversal wavefunction after traveling a distance in the
vacuum of length being smaller or equal than $10\fm$.
Accordingly, the jets should, to a large fraction,
materialize into hadrons still inside the expanding fireball, which has
a transversal size of roughly
$$
R(\tau ) \, \approx \, 8\fm +0.5\cdot \tau \, ,
$$
where $\tau $ denotes the local proper time of the
longitudinally (Bjorken like) and transversally expanding system
with velocity $v_t \approx 0.5 c$.

It is clear, though, that the parton cannot materialize into hadrons
if it still resides in a deconfined phase. The dressing with a
potential wavefunction and its subsequent fragmentation
can only occur outside of deconfined matter, i.e.~either
in the hadronic system or in vacuum. The physics of hadronization
out of a deconfined plasma for fast partons is not understood.
Thus the estimate (\ref{formationtime}) should really be seen
as a crude guess. The major fact we want to stress is that
the final energetic hadrons
(with a moderately high transverse momentum $p_\perp$)
observed at RHIC could well be realized
as nearly fully established hadrons already
inside the late stage hadronic fireball.
If this is the case, these (pre--)hadrons will interact
(by collisions) with the bulk hadronic matter making up the
fireball \cite{GGX02}.
In addition, refering to the phenomenology of colour transparency, 
the colourless objects becoming leading hadrons might interact as
pre--mesons 
with the surrounding low momentum mesons with a cross section of
roughly $\sigma (t) = \sigma_0\cdot(t/t_F)$ as long as
their wavefuction is not completely being built up \cite{DKMT}.
This picture of potential late hadronic final state interaction bears
again a close similarity to the one applied to explain
the $J/\Psi $ suppression pattern by either hadronic comovers
\cite{gavin} or by early hadronic string--like excitations as
early comovers \cite{jochen}.

The typical total (elastic plus inelastic) cross section of two mesons
is given by
$\sigma_0 \approx 10\dots20\mb $.
Indeed, at lower energies for the collisions, there might be strong
resonance contributions, which give rise to a much larger cross section.
The density of hadrons
in the late fireball changes from about $1\fm^{-3}$ down to
$0.1\fm^{-3}$. 
The mean free path of the 'fast' hadron is then estimated to be $\lambda
\approx 1\dots10\fm$. Accordingly, a few collisions $L/\lambda = 0,1,2,3
\ldots$ should take place. The system is potentially rather opaque!
In return, a strong degradation of the momenta of the leading
hadrons will result. This states the major motivation for the present
study: What is the potential influence of a specified finite number of
undergone collisions $L/\lambda = 0,1,2,3\ldots$ of the high momentum
particles on their spectrum? 

\begin{figure}[htb!]
  \begin{center}
    \includegraphics[width=6cm,angle=-90]{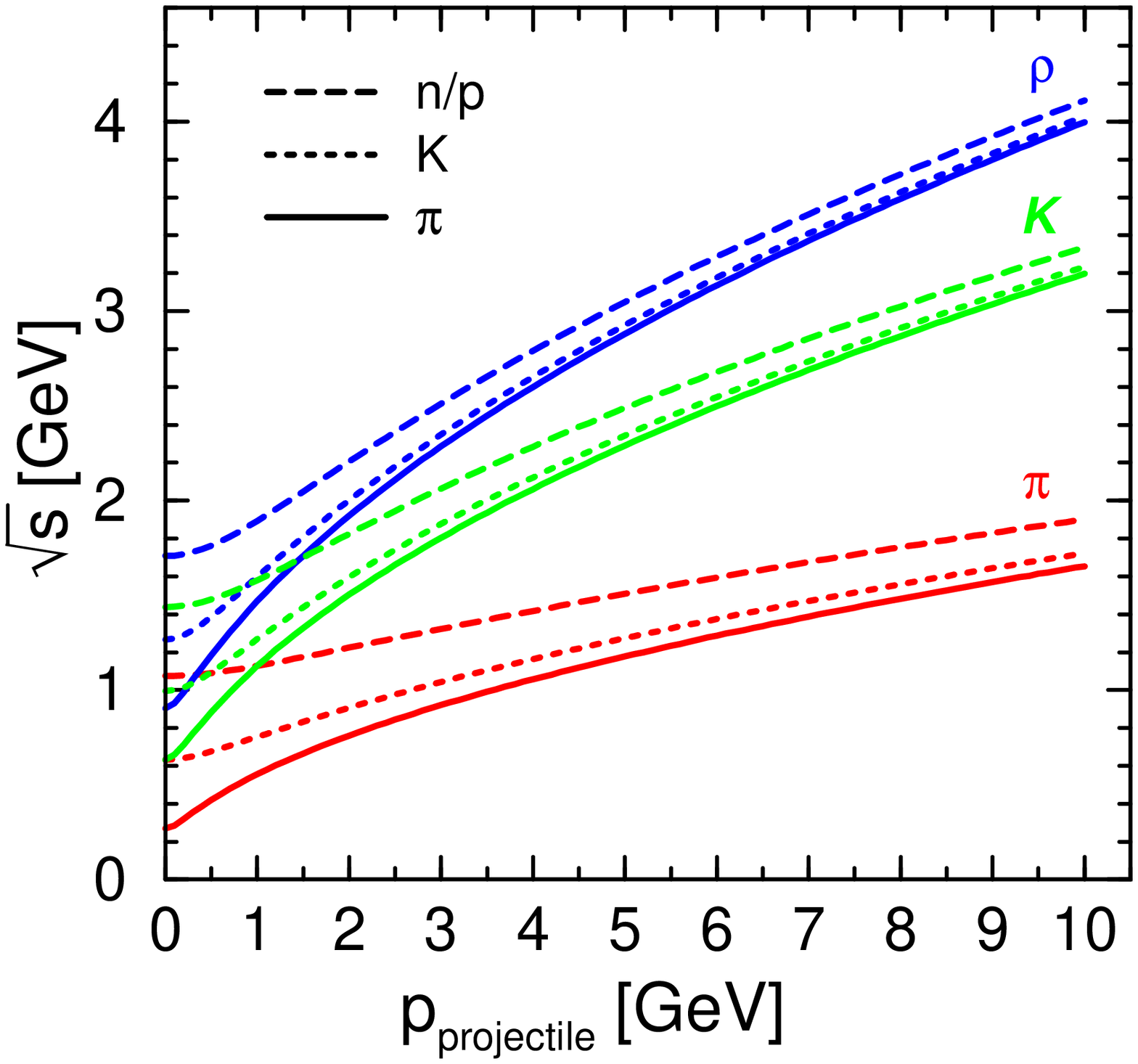}
    
    \CAPTION{
      The CMS energy for three targets at rest
      ($\pi,K,\rho$, depicted by the labels near the curves) and
      three types of projectiles ($\pi,K,n/p$, indicated by the solid,
      dashed and long dashed lines) as function of the
      momentum of the projectile.
      }
    \label{fig:SQRT}
  \end{center}
\end{figure}

Looking at the available center of mass energy for an individual collision
for mesons at the same rapidity,
even for a value of $p_\perp=10\GeV$, one gets a $\sqrt s <2\GeV$, if
the target is a pion
(assumed being locally) at rest. For a $\rho $ as target, one
has an invariant mass of above 2.5\GeV{} for particles with a
transverse momentum larger than 3\GeV{}. 
This behaviour is depicted in Fig.~\ref{fig:SQRT}.
In any case, at such lower energies the dominant part of the
collisions of the mesons or premesons with hadrons of the bulk system
is nonperturbative and cannot be described by pQCD methods.

For the considered transverse momentum region, one has either elastic
scattering, resonance scattering or also inelastic scattering
resulting in a few final hadrons. 
In section \ref{model} we explain the description of the collision
and introduce the notations of a folding matrix.
Since the surrounding hadronic gas is expected to consist of
approximately 40\% (direct) pions, 10\% kaons and 40\% rhos, we treat
in the following a typical inelastic collision within the 
\Fritiof{} scheme \cite{Fritiof} with a $\rho$--meson as characteristic
target hadron being at rest. 
To also stress some model dependence, we compare
with elastic scattering on a $\pi $ as target particle, where the
scattering is taken simply as isotropic.
In section \ref{results} we show various results
for the transverse momentum spectra at midrapidity for
various hadronic particles by assuming a definite
number of undergone collisions. The initial spectra are hereby
generated by the \Pythia{} code.
We end our discussion in section IV with a summary and some conclusions.


\section{Energy loss by (multiple) final state hadronic interactions}
\label{model}

Matter of interest are the transverse momentum spectra
$dN_i/dy\,d\vec p_\perp$, where the index $i$ stands for the various hadron
species and the transverse momentum vector $\vec p_\perp$ has to be
taken as two dimensional. 
In all what follows, we will assume a Bjorken--like
space--time picture with azimuthal symmetry, so that the only changes
of the spectra from the two dimensional transverse momentum vector
are given by its length $p_\perp$ and we therefore will skip for reasons
of simplicity all dependencies on the azimuthal angle.

In the framework of this assumption we are interested
in particles lying in the same rapidity window, introducing the
short hand notation $f_i(p_\perp)\equiv dN_i/dp_\perp$ for this
distribution function. (Our notation is different from
the more common form $dN_i/p_\perp dp_\perp$, which is due to
practical reasons in the descriptions presented below.)
In order to separate these distributions of (hard) initial
hadronic particles of type $i$
from the dominant distributions of
hadronic particles
of type $X$ with moderate or low (i.e.~`soft') momenta, say
with $p_\perp \leq 2$ GeV,
we also do introduce the distribution functions
$\tilde f_X(p_\perp,\phi)$. Here, $\phi$
does not reflect the azimuthal angle in the cylindrical shaped
Bjorken picture, but denotes the angle between the radial transverse
direction and its actual momentum orientation.
These latter distributions do describe the motion of the
target particles. They are assumed to be steeply falling functions in
the variable $p_\perp$, corresponding to a rather low temperature in a picture
of `thermalized' matter, superimposed by some collective, transversal
(`hydrodynamical') flow.

Within these definitions, one can express the resulting spectrum of
particles of the kind $j$ after one collision of a (hard) particle of
kind $i$ with a (soft) target hadron of kind $X$ as
\begin{eqnarray}
  f_j(p_\perp) &=& 
  \sum_i
  \int d\pT0\,f_i^0(\pT0)
  \ \int d\pT{X}\,d\phi_{_X}\label{eq:folding0}\\
  &&\quad\times 
  \frac{1}{\tilde N_X}
  \tilde f_X(\pT{X},\phi_{_X})\
  g^{X}_{ij}(\pT0;\pT{X},\phi_{_X};p_\perp)\ .
  \nonumber
\end{eqnarray}
Here $f_i^0(\pT0)$ has to be interpreted as the initial distribution
of particle kind $i$,
i.e.~the distribution one would assume without any modifications.
$\tilde N_X$ denotes the number of soft particles of kind $X$.
The ``folding matrix'' $g^X_{ij}(\pT0;\pT{X},\phi_{_X};p_\perp)$
expresses the probability to get a hadron of species $j$ with momentum
$p_\perp$, if a hadron of the species $i$ and the transverse momentum
$\pT0$ collides with a target hadron $X$, which is described by the
transverse momentum component $\pT{X}$ and the orientation
$\phi_{_X}$. 
Given this general structure with
four independent variables, $g^{X}_{ij}$ is a quite complicated
object.
It depends on the model employed for the collision (see below), and
which is connected with the target, as indicated by the upper index.

Since the non--transversal component of the target
momentum, which is encoded in the variable $\phi_{_X}$, also leads to
scattering in the longitudinal (beam) direction, equation
(\ref{eq:folding0}) 
is, strictly speaking, only true if one assumes that scattering--out
of a specified (tiny) region of rapidity is compensated by scattering--in
from the next rapidity regions.
This is the situation for the here assumed boost invariant Bjorken
picture of 1--dimensional longitudinal expansion.

Setting in the following
$\tilde f_X(\pT{X},\phi_{_X})/\tilde N_X=\delta(\pT{X})\,\delta(\phi_{_X})$,
we assume for a general estimate
that the low transverse momentum target
is approximated by $p_\perp\approx 0$, which is reasonable, as long as
the hard momentum scale is clearly separated from the
low momenta of the bulk hadronic particles.
This leads to the more simple and self--explanatory
folding equation for momentum degradation
\begin{equation}
  f_j(p_\perp) = \sum_i\int d\pT0\ f^0_i(\pT0)\ g^X_{ij}(\pT0,p_\perp)\ .
  \label{eq:model0}
\end{equation}
The basic problem is then effectively given by
one--dimensional collisions. Here the reduced folding matrix
$g^X_{ij}(\pT0,p_\perp)$ has a straightforward
interpretation: It indicates the probability that out of a given particle
$i$  with transverse momentum $\pT0$ one gets a particle $j$ with
transverse momentum $p_\perp$, if the collision is taken with a
target $X$ at rest, where all this happens in the same rapidity interval.
E.g. for elastic collisions $ i+X\rightarrow i+X $
($i\neq X$), one has the immediate
normalisation $\int_0^{\infty } d\pT \, g(\pT0,p_\perp)=1$.

Multiple, subsequent collisions are described by applying iteratively
eq.~(\ref{eq:model0}), where in a
new iteration step the distributions $f_i^0$ have to be replaced by
the $f_i$ resulting from the former step.

We do remark some words of caution:
\\
(1)
The separation of the transversal distribution into a ``hard''
contribution $f_i$ and a ``soft'' part $\tilde f_X$ and setting the
latter to a delta distribution implies some drawback,
as this separation is artificial at low $p_\perp$ (in the detector
one would see $f_i+\tilde f_i$) and thus leads to a ``particle
production'' effect at low values of $p_\perp$.
This is connected with the normalization of the folding matrix
$g^X_{ij}$, what one can see easily, if one for example looks at the
simple situation of elastic scattering $\pi^+$ on a $\pi$, i.e.~$i=\pi^+$ and
$X=\pi\equiv\frac13(\pi^++\pi^0+\pi^-$), leading to two pions with
non vanishing transverse momenta.
We remark here that one has to clearly stay
in the discussion above a certain
moderately large transversal momentum scale for an assumed
clear separation of hard and low momentum scales.
\\
(2) All our considerations assume, that a fixed number of collisions takes
place, disregarding the fact, that the cross section could vary
dramatically with the transverse momentum. For special collision
partners, for example $p+\pi$, it is known \cite{PDG}, that the cross
section $\sigma_{\rm tot}\left(p+\left[\pi^++\pi^-\right]/2\right)$ 
changes from $\approx30\mb$ for $p_{\rm proton}=3\dots6\GeV$ to nearly
five times this value at $p_{\rm proton}\approx1.8\GeV$, if the pion
is considerd to be at rest. 
Similarly, the elastic reaction $\pi+\pi\to\rho\to\pi+\pi$ also
exceeds $100\mb$ for the projectile momentum $p_\pi\approx 2\dots
2.5\GeV$.
This implies that, while assuming that
one collision appears, at an other value of the transverse momentum
(in these special cases) five collisons are to be awaited.
We will comment on this point a little further in the conclusions.


\subsection{Initial distribution}

For the initial distribution $f_i^0$ of individual hadrons $i$ at
moderately high $p_\perp$ when entering the final hadronic stage
we do not have direct information. All various ideas, as mentioned
in the introduction, might contribute and already steepen
the spectrum. In order to see clearly the potential effect
of the late hadronic interactions, we use individual
distributions generated by \Pythia{}
v6.2 \cite{PYTHIA} gauged to individual $pp$--collisions.

\begin{figure}[htb!]
  \begin{center}
    \includegraphics[width=6cm,angle=-90]{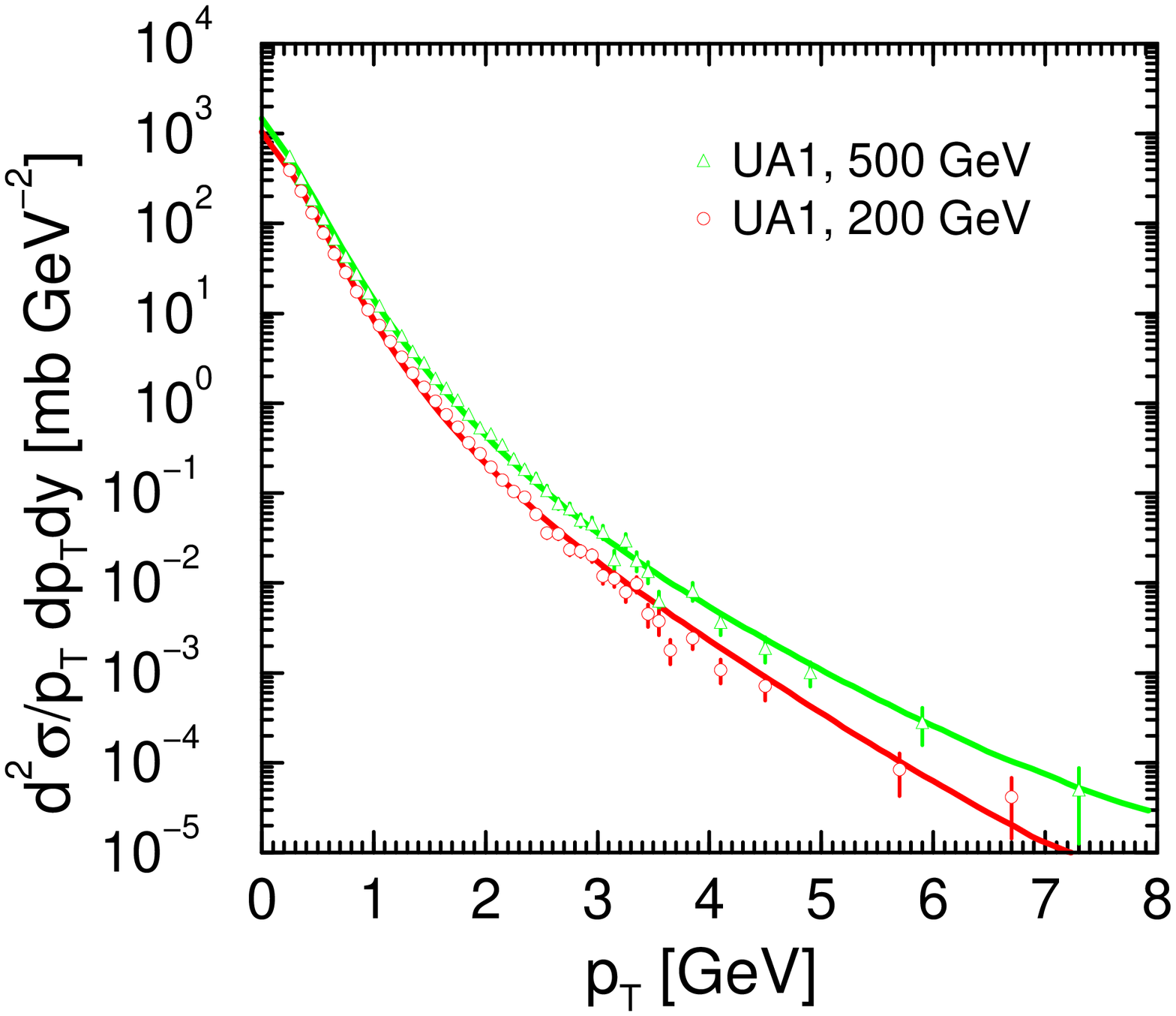}
    
    \CAPTION{
      Charged hadron distributions for $p\bar p$ collisions
      calculated with \Pythia{} at $\sqrt s = 200,\,500\GeV$ 
      compared with experimental data \cite{Albajar90}.
      }
    \label{fig:ppbarDistr}
  \end{center}
  \begin{center}
    \includegraphics[width=5.5cm,height=3.816cm,angle=-90]{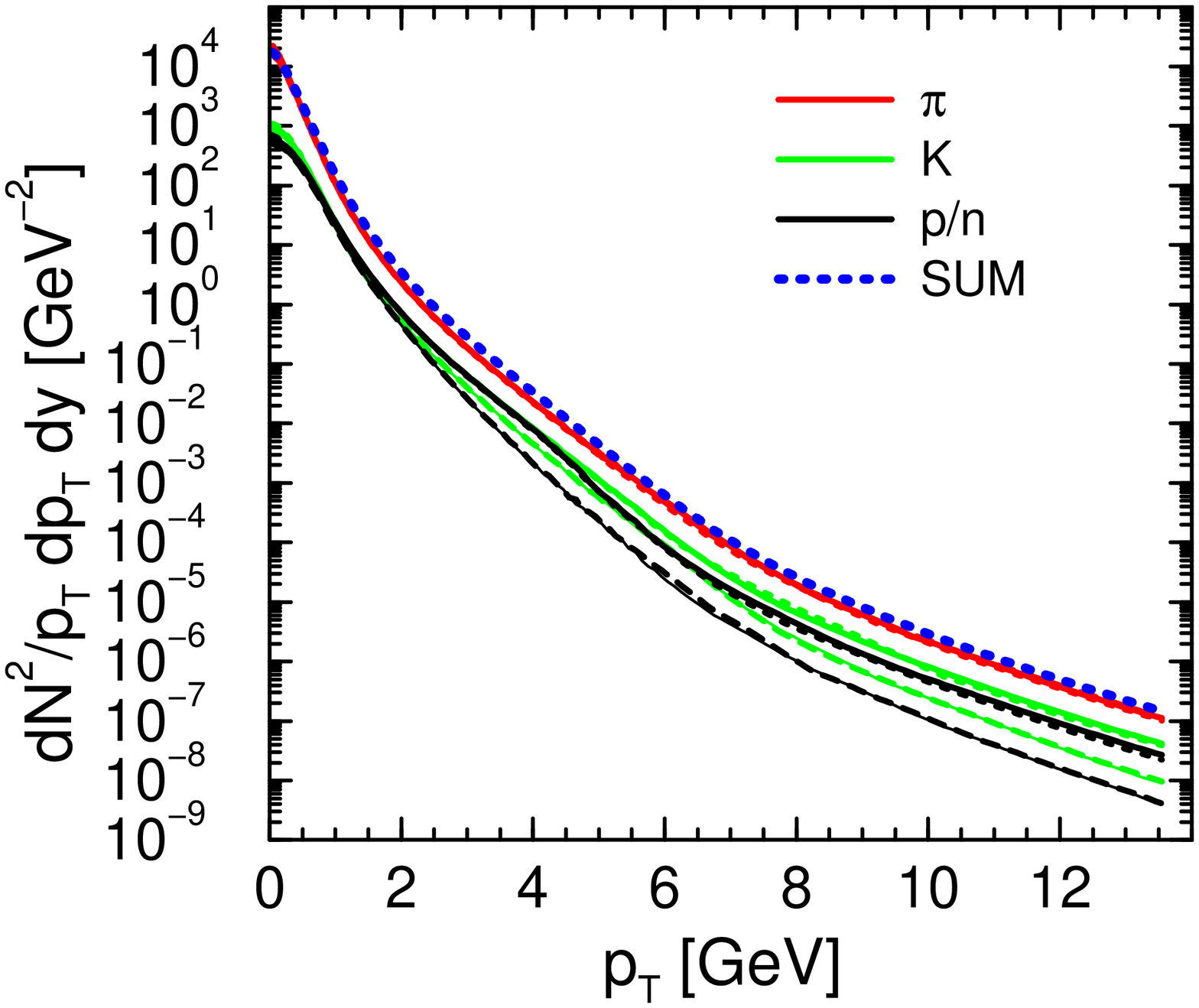}
    \hspace*{-0.15cm}%
    \includegraphics[width=5.5cm,height=3.184cm,angle=-90]{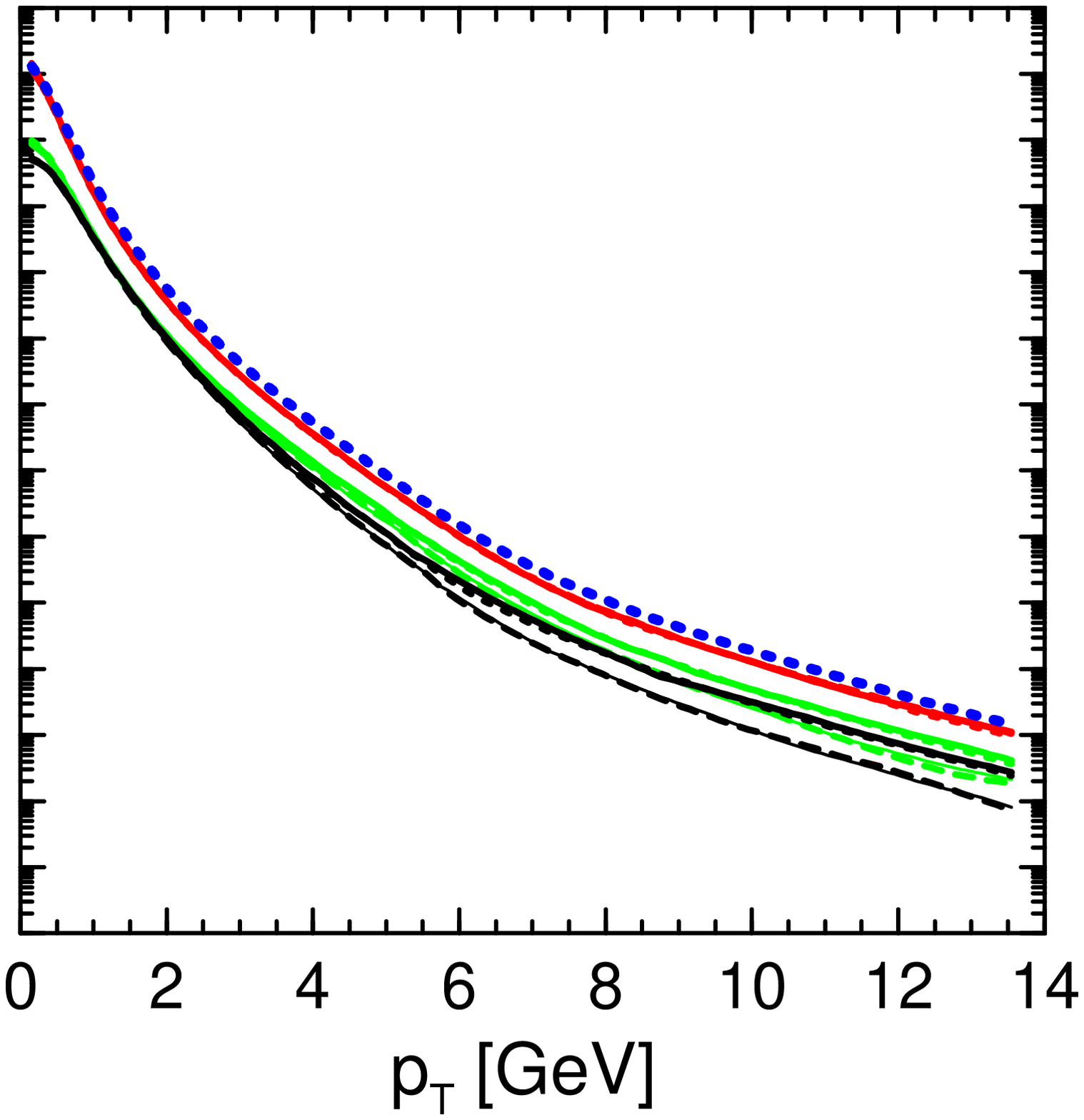}
    
    \CAPTION{
      Hadron distributions at midrapidity for $Au$--$Au$ collisions
      at $\sqrt s = 130\GeV$ (left) and $\sqrt s = 200\GeV$ (right)
      calculated with \Pythia{}.
      Pions, kaons and protons/neutrons
      are seperated by line colour, while their charge state
      [negative (long dashed), neutral (solid), positive (dashed)] is
      indicated by the line style. Neutral anti--particles are
      displayed with a thin line.
      The sum of all charged particles is depicted by the thick
      dotted curve.
      Note that the spectra of all pions are similar and the spectra of
      neutral (anti)kaons and nucleons are similar to that of their
      positive (negative) charged partners.
      }
    \label{fig:AuAuDistr}
  \end{center}
\end{figure}

It is essential to carefully adjust the parameters for the model
calculations to data provided by the UA1 collaboration for $p\bar p$
collisions in the energy range $\sqrt s \geq 200\GeV$ and $|y|<2.5$
\cite{Albajar90}.
Fitting the calculations to the data, one gets a best
description of the experimental results by just tuning one parameter,
the intrinsic transverse momentum distribution of the partons inside
the nucleon, described by its averaged value $\kTave{} (\sqrt{s})$.
The resulting spectra are displayed in Fig.~\ref{fig:ppbarDistr}.

In a second step one has to compare the calculated spectra with the 
parametrization $(1+p_\perp/ \bar{p}_0)^{-n}$ used to describe
the experimental data \cite{Albajar90} and extract
from this fitted parmetrization
($\bar{p}_0  (\sqrt{s})$ and $n(\sqrt{s })$)
the characteristic value
$\pTave{}(\sqrt s)$.
With this knowledge one can extrapolate the calculations of 
$p\bar p$ collisions down to the energy  $\sqrt s=130\GeV$
by adjusting the intrinsic parameter $\kTave{} $ to finally
meet the experimentally extrapolated value
$\pTave{}(\sqrt s=130\GeV)=0.385\GeV$ \cite{Albajar90}.
Please note that this value is completely disregarded by the
assumed pointwise interpolation of fits to data from different
values of $\sqrt{s}$ for
given transverse momentum $p_\perp$, as used eg.~in 
\cite{PHENIX,Ex130}.

After having adjusted the microscopic
Monte--Carlo calculations for $p\bar p$
collisions at $\sqrt s=130$ and $200\GeV$ and $|y|<2.5$, theoretical 
predictions for $Au$--$Au$ collisions at these energies for
midrapidity are given by respecting the correct isospin average
corresponding to 
$0.16\cdot pp + 0.36\cdot nn + 0.48\cdot pn$. 
It is worth noting, that these distributions do not follow a true power
law behaviour, as can be seen in Fig.~\ref{fig:AuAuDistr}, where at
$p_\perp\simeq2\GeV$ and $\simeq6\GeV{}$ significant changes in the slope occur.

Some special care has to be taken (for example mixing of calculations with
non--weighted and some with reweighted events) to expand the \Pythia{}
Monte--Carlo calculations up to values of $p_\perp$ larger than $10\GeV$
with best statistics. 
Naive Monte--Carlo implies, that at $p_\perp\approx6\GeV$ reasonable
computation time ends and that one would have to increase the number of
events by approximately a factor of 10 every time one wants to go
further in the $p_\perp$ spectrum by an other step of 2\GeV, as one can
see in Fig.~\ref{fig:AuAuDistr}.


\subsection{Folding Matrices}

The collision of a high $p_\perp$ particle with particles of the medium is
implemented according eq.~(\ref{eq:model0}) via a
folding matrix $g_{ij}(\pT0,p_\perp)$.
Since a unique description of scatterings at these
low energies (cf.~Fig.~\ref{fig:SQRT}) is not available, model
descriptions have to be choosen.

In our first description A we calculate the folding matrices with
Monte--Carlo calculations using the package \Fritiof{}
\cite{Fritiof}
to simulate collisions with a
$\rho $-meson. 
Since it is known, that within \Fritiof{} the relative weighting
between elastic and inelastic scatterings is not in full agreement
with data, we therefore choose a constant weighting according the
ratio 15:85. 
(This is
given by the averaging of $p+\rho$--, $\pi+\pi$--, $K+\pi$ cross
sections in a pomeron/reggeon exchange picture for $\sqrt s>2\GeV$ and also
following simple estimates in a constituent quark model.)

As a second description B we use purely elastic scattering on a pion,
where the angular distribution of the outgoing particles in the cm system is
assumed to be isotropic, which seems to be a at a first glance a
rather strong assumption.

In the description A of the collisions we are forced
to extrapolate the \Fritiof{} prescription downwards to energies,
where it is strictly not fully applicable anymore (for values below
an initial transverse momentum of 
$\pT0\simeq3\GeV$ we are in the resonance region,
cf.~Fig.~\ref{fig:SQRT}).
Therefore (or nevertheless) we stop the calculations at a
minimal initial transverse momentum of about $\pT0^{\rm min}=1\GeV$.
Boundary problems applying the folding procedure acccording 
eq.~(\ref{eq:model0}) 
due to this cut could be easily cured,
if one implements a more detailed weighting of inelastic and elastic
cross sections, which could already allow to bend the calculations
downwards to zero transverse momentum, or if one would be able to
introduce a consistent picture of scattering of hadrons from very low
(resonance region) to very high momenta (hard QCD processes).

\begin{figure}[htb!]
  \begin{center}
    \includegraphics[height=7.5cm,clip=true]{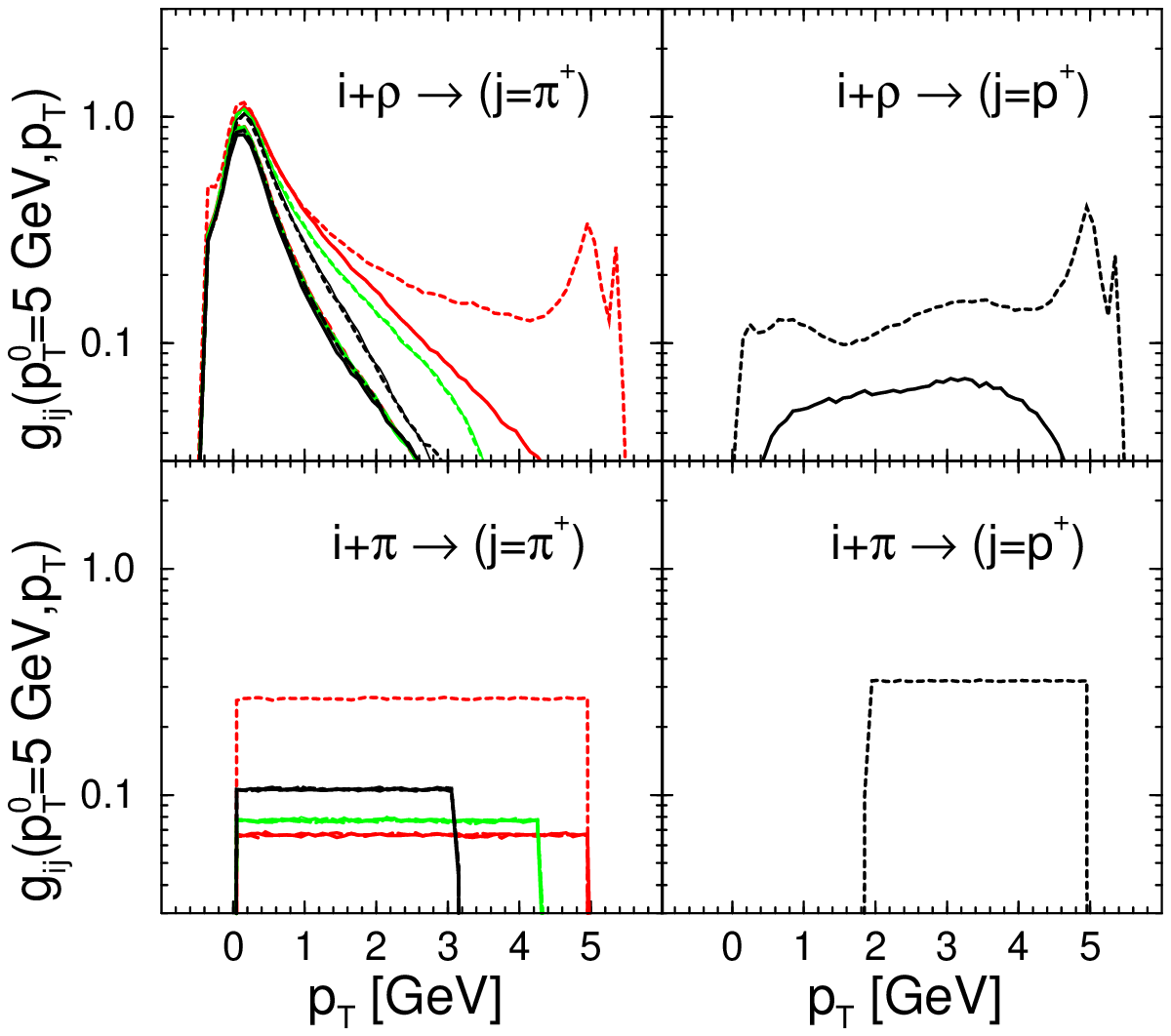}

    \CAPTION{
      The folding matrices $g$ according
      (in)elastic scattering on a $\rho$ (top) or elastic
      scattering on a $\pi$ (bottom) for a $\pi^+$ (left) and a $p$
      (right) as outgoing particle for a fixed value of the initial
      transverse momentum $\pT0=5\GeV$ as function of the resulting
      transverse momentum $p_\perp$. The line styles are choosen
      as in Fig.~\ref{fig:AuAuDistr}, indicating the different
      incoming particles.
      `Negative' values for transverse momenta $p_\perp$ stand for
      scattering in the opposite direction with positive
      momenta. 
      }
    \label{fig:g_pT}
  \end{center}
  \begin{center}
    \includegraphics[height=7.5cm,clip=true]{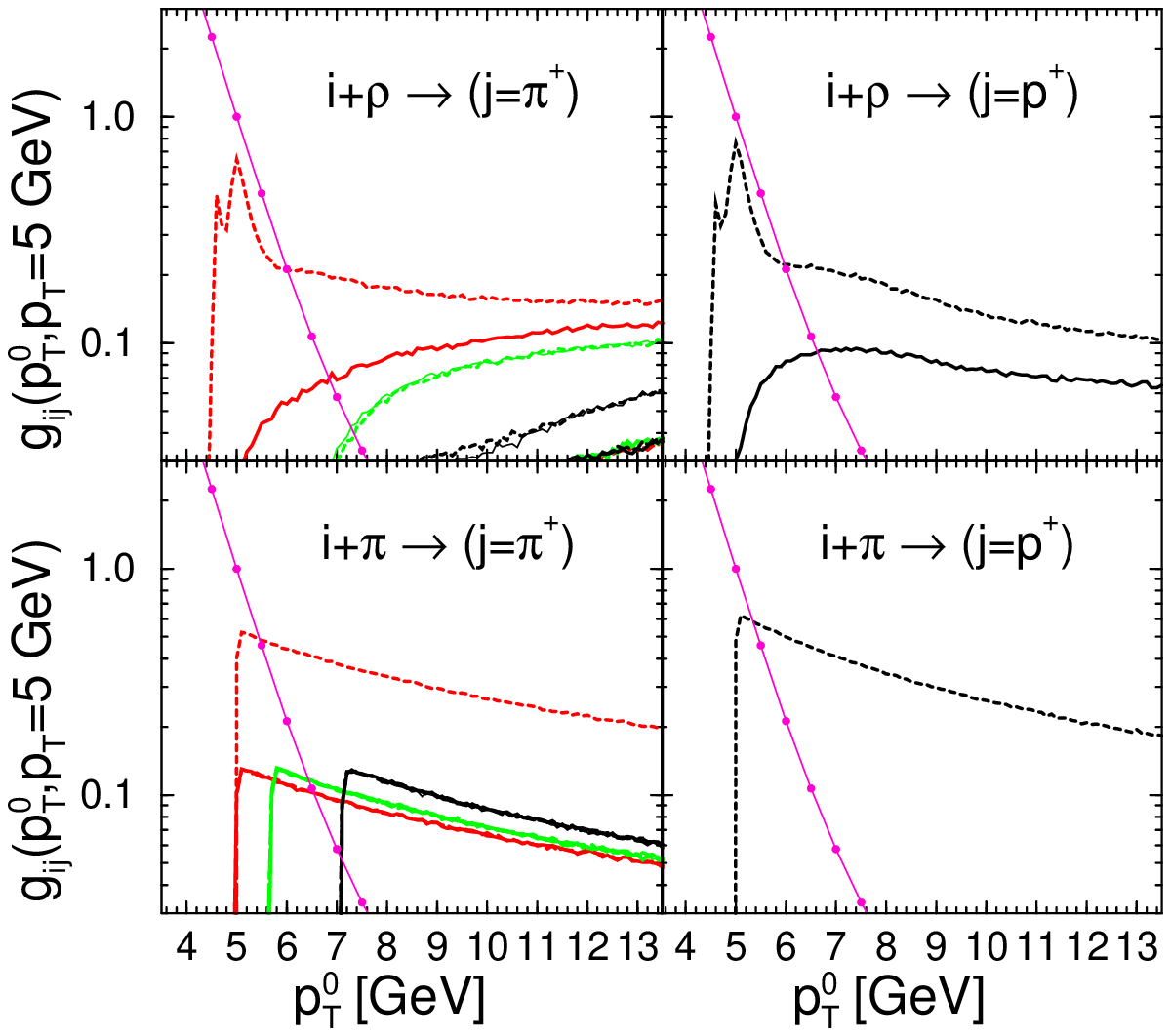}

    \CAPTION{
      As Fig.~\ref{fig:g_pT}, but now as function
      of the initial transverse momentum $\pT0$ for a given
      resulting transverse momentum $p_\perp=5\GeV$.
      The connected dots stand for the slope of
      $f_{(h^++h^-)/2}(\pT0)$ at ${\sqrt s=200\GeV}$.
      }
    \label{fig:g_pT0}
  \end{center}
\end{figure}

The two different model descriptions show some significant
differences, if one looks at the 
dependence of the final transverse momentum $p_\perp$ for a given
initial transverse momentum $\pT0$. 
But not only the different models, but also different final particles
behave differently. 
This is shown in Fig.~\ref{fig:g_pT}, where as an example the
convolution matrizes $g^{(\pi)}_{i\,\pi^+/p^+}$ and 
$g^{(\rho)}_{i\,\pi^+/p^+}$ are shown for the model descritpions A and
B for the fixed value of the initial transverse momentum
$\pT0=5\GeV$.

While all the elastic distributions (Fig.~\ref{fig:g_pT} (bottom))
do follow a flat behaviour, where the 
threshold values are dictated by
$\pT0$ and the corresponding masses of the particles,
the inelastic distributions (Fig.~\ref{fig:g_pT} (top))
do possess a much richer structure.
First they show, as expected,
a strong enhancement at low $p_\perp$ values.
In addition, these curves also depict
a double peak structure for the largest $p_\perp$--values
in the forward scattering region. The lower peak at
$p_\perp=\pT0$ is the simple one due to elastic scattering, while the second
and rather astonishing peak for $p_\perp>\pT0$ is due to a large partial
cross section for $X+\rho\to X+\pi$
implemented in \Fritiof{} ($X$ denotes here either the $\pi^+ $
in the upper left figure or the proton in the upper right one
of Fig.~\ref{fig:g_pT}). 
This peak is located at
$(p_\perp-\pT0)\simeq(m_\rho^2-m_\pi^2)(2m_\rho)\approx m_\rho/2$.
Physically, for such collision events
the leading particles do become slightly accelerated.
One might argue, that the likelihood of such contributions
are completely overestimated, so that \Fritiof{}
should also fail in this part of the description. 
On the other hand, by
including them, their net effect in
our later discussion would be that one actually
underestimates the emerging depletion of the transverse momentum spectrum.
We have decided to omit these contributions from the inelastic part by
hand compared to the basic \Fritiof{} concept.

Considering eq.~(\ref{eq:model0}), one
realizes the fact that the behaviour of the folding
matrices for a fixed value of the resulting transverse momentum
$p_\perp$ as a function of the initial transverse momentum $\pT0$
is of crucial importance. The corresponding figures for
the examples from above are depicted  in Fig.~\ref{fig:g_pT0}.

While the curves for the elastic scattering on a pion 
are smoothly falling for $\pT0\geq p_\perp$ (or the corresponding threshold
value), the curves for the scattering on a 
$\rho$ are nearly constant, but also showing the double peak structure
at $p_\perp=\pT0$ and  $p_\perp>\pT0$, as
mentioned above. For larger $\pT0$ the behaviour of all these
folding curves gets less and less important for the final
spectrum, since one has to fold them with the ininital
distribution, whose steep slope is also indicated in
Fig.~\ref{fig:g_pT0}.
This means that the correct way of looking at the stopping is the
function $g_{ij}(p_\perp=const)$ instead of
$g_{ij}(\pT0=const)$. 
Looking at the potential size of the energy loss
(at a partonic or a hadronic level),
the value $\pT0 - p_\perp $ of $g_{ij}(\pT0,p_\perp)$
is strongly skewed toward small values
by the steeply falling distribution $f_i^0(\pT0 )$.
Only the least
stopped particles (i.e.~particles with $\pT0$ only somewhat larger
than $p_\perp$) in $g_{ij}$ contribute to the folding integral
eq.~(\ref{eq:model0}).
Simple estimates give, that values of $g_{ij}$ with
$p_\perp<\frac12\pT0$ contribute only at a one percent level.
Although appealing from the various shapes of the
folding matrix $g_{ij}$, assumptions like
$\Delta E = \Delta p_\perp = \left<p_\perp-\pT0\right> \simeq
const\cdot\pT0$ for an averaged energy loss
are much too oversimplistic.

These findings are in complete agreement with the
considerations of \cite{Baier},
where the energy loss of (very energetic) partons in partonic matter
due to gluon 
radiation, calculated via perturbative QCD, is expressed via a
function $D(\epsilon)$, which has to be integrated in a sensible way
with $\epsilon$ as a measure of the energy shift.

Inspecting further Fig.~\ref{fig:g_pT0},
the $p_\perp>\pT0$ peak of the \Fritiof{} results will lead to some 
(maybe unphysical) ``unstopping'', as already emphasized.
As those contributions are indeed then weighted with the much larger
distribution $f(\pT0 )$ at $\pT0 < p_\perp $, they would indeed be
of more crucial importance than naively expected.
Including those contributions would roughly weaken the deacceleration
effect in the spectrum by a factor of 2 for $p_\perp\geq3\GeV$ and lead
to a zero net effect at $p_\perp\simeq2\GeV$.
As already discussed, we will exclude these contributions with  
$\pT0 < p_\perp $.


\section{Quenching by multiple collisions and comparison with data}
\label{results}

\begin{figure}[htb!]
  \begin{center}
    \includegraphics[width=6cm,angle=-90]{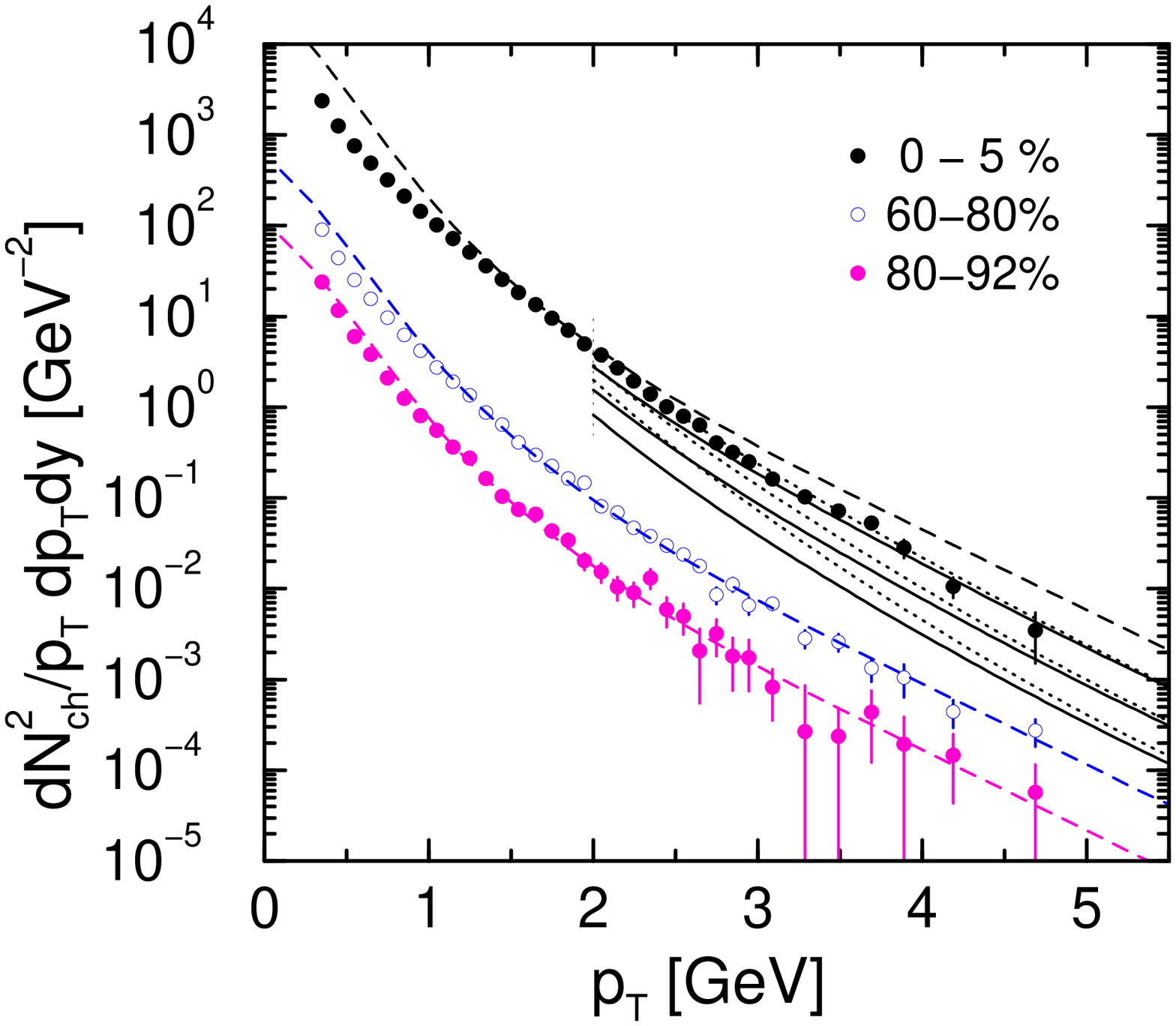}
    \CAPTION{
      Comparison of the calculations at midrapidity with the
      (at that time preliminary) PHENIX data \protect\cite{PHENIX2}
      (but see also \protect\cite{PHENIX})
      on charged hadrons for three different centralities.
      The thin dashed lines indicate the results from \Pythia{}
      scaled with a corresponding binary collision number
      $N_{coll}$.
      The solid lines depicts the spectra for the
      most central region, where the particles have suffered on
      average $\langle L/\lambda\rangle\equiv 1, \, 2, \, 3$
      (in)elastic hadronic collisions on a $\rho $,
      while the dotted lines indicate the same effect for elastic
      scattering on a $\pi$.
      These modifications are only shown for $p_\perp>2\GeV$.
      }
    \label{fig:PH_ALL}
  \end{center}
  \begin{center}
    \includegraphics[width=6cm,angle=-90]{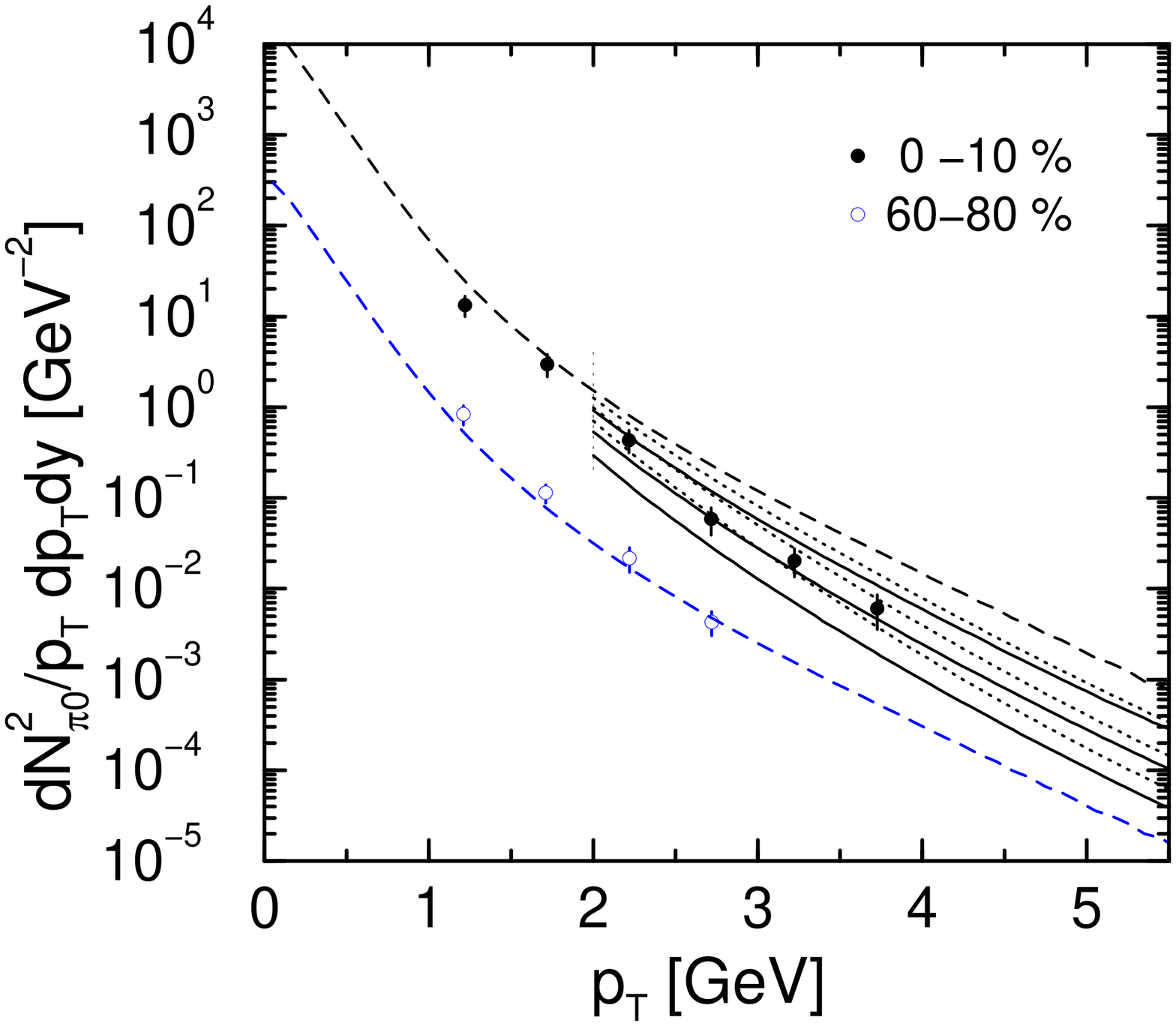}
    \CAPTION{
      As Fig.~\ref{fig:PH_ALL}, but here a comparison of the
      present calculations with the PHENIX data
      \protect\cite{PHENIX2} on $\pi _0$ is depicted.
      }
    \label{fig:PH_PI0}
  \end{center}
\end{figure}

Employing the appropriate scaling of binary collisions
($N_{coll}=945,\,905,\,19,\,3.7$ for the centralities 0\dots5\%,
0\dots10\%, 60\dots80\% and 80\dots92\% respectively),
the \Pythia{} calculations are in perfect agreement with the
(at that time preliminary) PHENIX data \cite{PHENIX2}
for charged hadrons and for neutral pions in the case of
peripheral collisions, as can be seen in Figs~\ref{fig:PH_ALL} and
\ref{fig:PH_PI0}. We can thus conclude that our \Pythia{} calculation
with the correct extrapolation provides an accurate basis for the
experimental comparison.

Looking at the most central data, one recognizes the celebrated
significant discrepancy between
our scaled \Pythia{} results and the data.
The data resides below the
theoretical curve for $p_\perp<1.5\GeV$ and for $p_\perp>2.5\GeV$.
Only the latter case for $p_\perp \geq 2 \GeV $
we now want to adress by means of our
scenario of possible (multiple) collisions of the
jet--like (pre--)hadrons with the `soft' hadrons sitting in the
bulk matter of the fireball.
There is some general agreement that the hadron spectra
below $p_\perp \simeq 2\GeV$ should be dominated by soft physics and
hydrodynamical expansion with some appreciable collective outward flow.
In addition, in this lower momentum regime our basic
assumption of a hadronic target being approximately at rest
becomes no more valid.
Here not only energy loss, but probably also some collisional energy gain
(on a partonic and/or hadronic level)
and also hadronic particle annihilation
and (re--)absorption occurs which influences
the shape of the spectrum.

Refering to eq.~(\ref{eq:model0}) and inspecting the two
Figs.~\ref{fig:g_pT} and \ref{fig:g_pT0}, one recognizes that the
depletion in the spectrum by one collision should be stronger
for description A with (in)elastic collision on a $ \rho $ as the soft
target particle, as one would also intuitively expect. 
This is so as the foldings matrices $g_{ij}$ for $\pT0>p_\perp$ are 
somewhat smaller for this first collision picture, though there is the
same strength at $\pT0\simeq p_\perp$.
In Figs~\ref{fig:PH_ALL} and
\ref{fig:PH_PI0}
the three solid lines depicts the spectrum (for $p_\perp>2\GeV$) for the
most central region,
when the particles have suffered on
average 1,2,3 (in)elastic hadronic collisions on a $\rho $.
The dotted lines in these figures indicate the effect for elastic
scattering on a $\pi$.
Indeed, for the inelastic case the overall depletion is somewhat
stronger.
Refering to our earlier remarks,
we only do show the modifications for $p_\perp>2\GeV $.
One clearly recognizes the potential importance of
these possible final state interactions.
For $p_\perp>2\GeV$ an average loss
with $\langle L/\lambda\rangle =1\cdots2$ collisions
are appropriate to
explain the present experimental data both for charged
hadrons and neutral pions.

We remark, that the potential energy loss by final state hadronic reactions
is in the same range as considered in the various discussions
for the jets in possible deconfined matter.

\begin{figure}[htb!]
  \begin{center}
    \includegraphics[width=5.5cm,angle=-90]{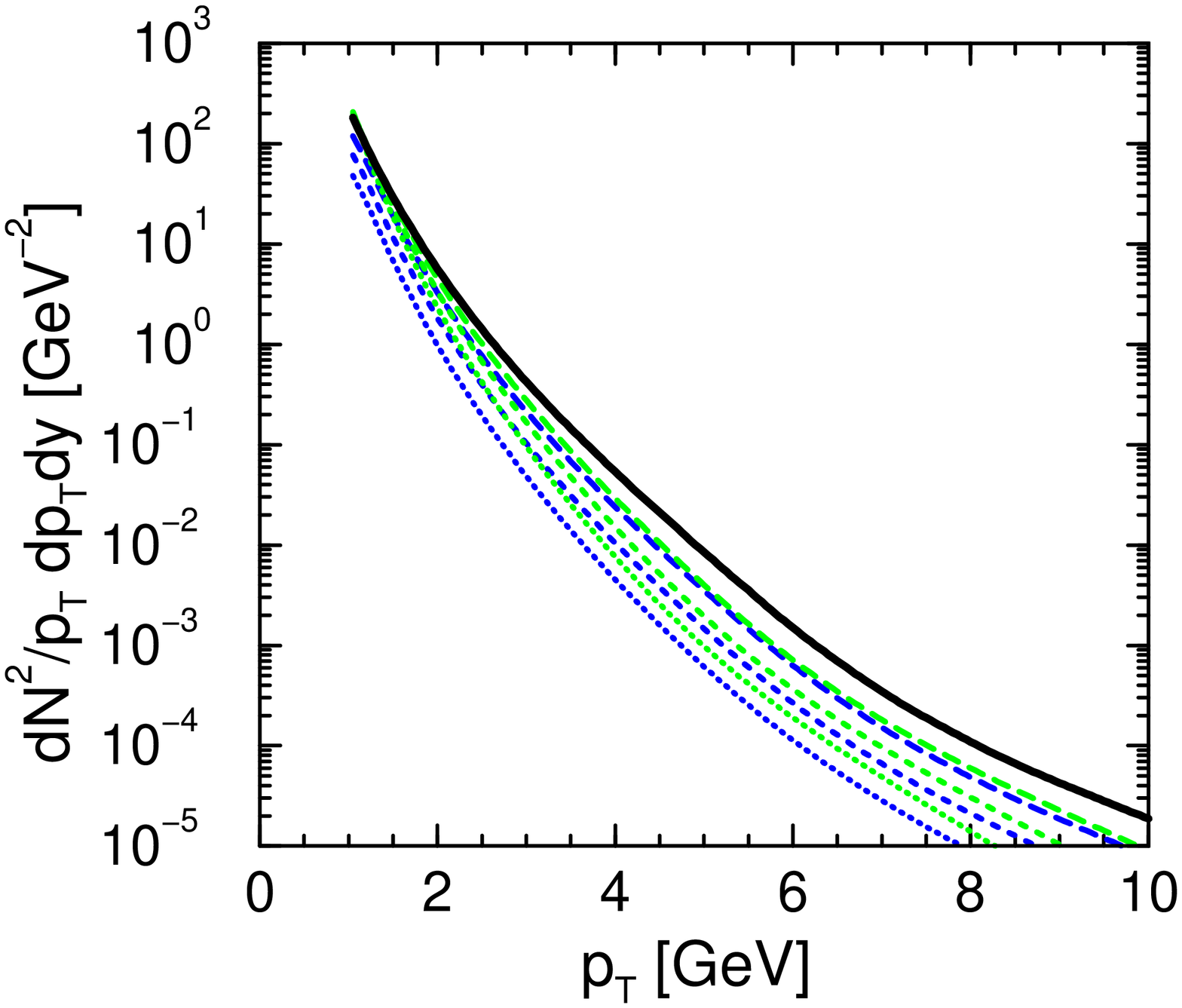}
    \CAPTION{
      Resulting $p_\perp$--spectra of charged hadrons at midrapidity
      for $\sqrt{s}=200\GeV$ (solid black line) and for
      $\langle L/\lambda\rangle\equiv 1, \, 2, \, 3$ (top to bottom) 
      collisions according (in)elastic scattering on a $\rho$ (blue)
      or elastic scattering on a $\pi$ (green-light).
      }
    \label{fig:PH_PRED}
  \end{center}
  \begin{center}
    \includegraphics[width=5.5cm,angle=-90]{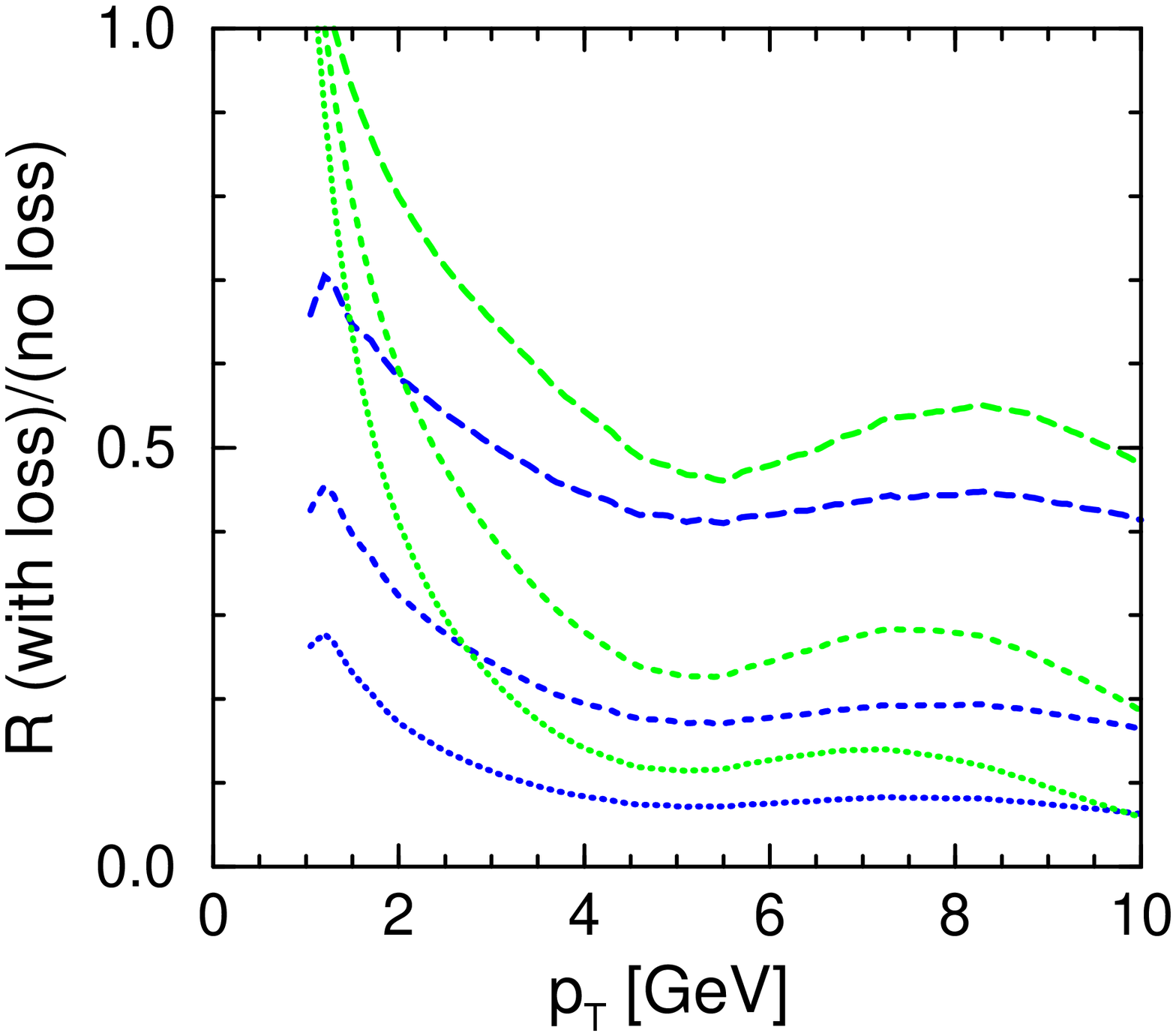}
    \CAPTION{
      The suppression factor $R(p_\perp)$ of charged hadrons at
      midrapidity for $\sqrt{s}=200\GeV$ for
      $\langle L/\lambda\rangle\equiv 1, \, 2, \, 3$ (top to bottom) 
      collisions according (in)elastic scattering on a $\rho$ (blue)
      or elastic scattering on a $\pi$ (green).
      }
    \label{fig:PH_PRED_R}
  \end{center}
\end{figure}

In Figs.~\ref{fig:PH_PRED} and \ref{fig:PH_PRED_R} we now show
calculations
for various number of collisions 
at a $\sqrt{s_{NN}}=200 $ GeV up to $p_\perp = 10 $ GeV.
Our results might serve also as a guideline and prediction
concerning the present analysis of the 200 GeV runs at RHIC \cite{QM2002}.
In principle, the results are completely similar to the one carried
out at the lower RHIC energies. 
Fig.~\ref{fig:PH_PRED} shows the resulting $p_\perp$ spectra after
1,2,3 collisions according to the two collision pictures.
As expected, the modification of the spectrum is similar to the one
discussed for the situation at lower energies of $\sqrt s=130\AGeV$.
Thus we can conclude that hadronic final state collisions
are efficient in degrading the transverse energy of jet--like hadronic
particles. 

Fig.~\ref{fig:PH_PRED_R} gives the resulting suppression factor
$R(p_\perp)
=\frac{\left.dN/dp_\perp\right|_{\text{modified}}}
{\left.dN/dp_\perp\right|_{\text{initial}}}$ 
for collisions from Fig.~\ref{fig:PH_PRED}.
Here the effect of the different collision schemes is more visible.
With one to two inelastic collisions one will have a suppression of
50 to 80 percent for momenta $p_\perp \geq 4 \GeV $.


\section{Summary}

We have motivated that
most of the (pre--)hadrons stemming
from a jet should still materialize in the dense system
for transverse momenta up to 10 GeV. The late hadronic
final state interactions with the bulk of comovers have
a clear and nonvanishing effect in suppressing the spectrum.
This is so for (in)elastic reactions on a $\rho$, which we have
modelled via the FRITIOF scheme, and also for (isotropic) elastic
scattering on a $\pi$.
On the average one up to two such interactions should already be enough
to explain quantitatively the RHIC results.

Although this comparison of the depletion of the spectra
with real data is intriguing, one should be aware that
our calculations are at best semi--quantitative. This
is so because of a couple of simplifying assumptions invoked
in the present study.
We have neglected all possible dependencies of the cross
sections of the colliding ``hard'' and ``soft'' (target) particles,
but do assume that a fixed number of collisions has taken place,
following a simple--minded opacity expansion.
In particular, possible resonance contribution for the elastic
scattering could enhance the effect of momentum degradation considerably.
Second we made the further assumption, that the bulk hadronic targets
are taken at rest. This is certainly not the case especially
if one wants to quantitatively discuss the modifications
in the spectra around $p_\perp= 1 \ldots 3 \GeV $, where both collective
and `thermal' dynamics should play a considerable role.
Energy gain and loss mechanisms should both be at work and
are probably competing in this momentum range.
On a partonic level such a picture has been discussed in \cite{Wang2},
but this should be rather obvious if one thinks of a kinetic cascade--type
description either on a partonic or hadronic description of the dynamics
happening inside the fireball.
Hence, the marriage of a transport description for the `soft' matter
by some standard ultrarelativistic heavy ion cascade
together with the present ideas carried out in this study
is one possible way to further proceed in order to obtain
more quantitative results. We will leave this as a future option.
A first and rather exploratory investigation in this direction
has already been mentioned some time ago \cite{Bass}.

Our emphasis is, however, at this stage to point out that,
in order to draw some deductions for possible QCD effects
of a deconfined QGP phase on the materializing jets, one has to
disentangle these from the here investigated final state interactions,
before definite conclusions on the importance of a potential dense
partonic phase, or any other effect, can
really quantitatively be drawn.
This is certainly true for the case of the jets to be observed at
RHIC.
(At future LHC experiments observable jets with much higher $p_\perp$
could actually materialize outside the fireball.)

Combining specific energy loss fomulae with fragmentation functions of
quarks and gluons into hadrons can therefore only be understood
as an extremely ``effective'' approach,
since no conclusive answer about the relative
strength of hadronic final state interactions can be extracted.
The possible disentanglement, though, is very challenging and
delicate.
One possible way could be the precise measurements of
$\pT{}$--distributions of charmed particles like the $D$--mesons
\cite{referee}.
According to a recent investigation \cite{DokKha01} heavy quarks
should loose considerable less energy than light quarks or gluons
(although an earlier study comes to other conclusions \cite{Munshi}).
On the other hand the (elastic and inelastic) $D$--meson hadronic
cross section should not be much smaller than that of light quark
mesons.
Hence, if the $D$--mesons should not show any significant
$\pT{}$--suppression at moderate $\pT{}\sim 3\dots10\GeV$, this would
then favour the partonic 'pQCD' jet quenching scenario. At present,
the PHENIX data \cite{PhenixEl} up to $\pT{}$ values of $3\GeV$ still
suffer from too large statistical errors and are compatible with full
or no quenching \cite{KaiQM}.
In any case, the $D$--meson spectra will be an important experimental
challenge for future RHIC undertakings.
Some other valuable information on the right ``initial distribution''
will be learned by $d+Au$ experiments \cite{VG02}, eg. the size of
possible shadowing and the Cronin effect. However, like
the medium--induced gluon radiation the here discussed final state
interactions with the late stage hadronic matter is a pure heavy ion
effect. Here the $p_\perp$--dependence of the `suppression' pattern
might reveal possible differences, as for higher $p_\perp$ the time of
formation 
for the hadronic wavefunction of the leading hadrons of a
jet and the colour transparency effect should suppress these final state
interactions. But this is speculation at present.

\vspace*{\fill}

\subsection*{Acknowledgments}

Stimulating discussions with R.~Baier, W.~Cassing, T.~Falter,
P.~Jacobs, B.~K\"ampfer, U.~Mosel and J.~Nagle are gratefully
acknowledged.
This work has been supported by BMBF.




\begin{thebibliography}{99}
\itemsep=0cm
\bibitem{PHENIX} K.~Adcox et al.~[PHENIX],
\Journal{\PRL}{88}{022301}{2002};
K.~Adcox et al.~[PHENIX],
{\em nucl-ex/0207009}.

\bibitem{STAR} 
C.~Adler et al.~[STAR],
\Journal{\PRL}{87}{112303}{2001};
C.~Adler et al.~[STAR],
{\em nucl-ex/0206011}.



\bibitem{QM2002} various presentations at the Quark Matter 2002 conference,
proceedings to be published in {\em Nucl.~Phys.} A.

\bibitem{Wang} J.D.~Bjorken, FERMILAB-PUB-82-59-THY (unpublished);
X.N.~Wang, M.~Gyulassy, \Journal{\PRL}{68}{1470}{1992}.

\bibitem{Wang1} X.N.~Wang, \Journal{\NPA}{698}{296}{2002}.

\bibitem{BDMS}
R.~Baier, Y.~Dokshitzer, A.~Mueller, S.~Peign{\'e}, D.~Schiff,
\Journal{\NPB}{483}{291}{1997};
\Journal{\NPB}{484}{265}{1997}.

\bibitem{Baier}
R.~Baier, Y.~Dokshitzer, A.~Mueller, D.~Schiff,
JHEP {\bf 109}, 033 (2001).

\bibitem{Kai} K.~Gallmeister, B.~K\"ampfer, O.~Pavlenko,
\Journal{\PRC}{66}{014908}{2002}.

\bibitem{Levai}
M.~Gyulassy, P.~Levai, I.~Vitev,
\Journal{\NPB}{571}{197}{2000};
\Journal{\NPB}{594}{371}{2001};

\bibitem{Levai2}
P.~Levai, G.~Papp, G.~Fai, M.~Gyulassy, G.~Barnaf\"oldi,
I.~Vitev, Y.~Zhang, \Journal{\NPA}{698}{631}{2002}.


\bibitem{JMS02} S.~Jeon, J.~Jalilian-Marian, I.~Sarcevic,
{\em hep-ph/0207120}.

\bibitem{VG02} I.~Vitev, M.~Gyulassy,
{\em hep-ph/0209161}.

\bibitem{SKMV01} J.~Schaffner-Bielich, D.~Kharzeev, L.~McLerran,
R.~Venugopalan, \Journal{\NPA}{705}{494}{2002}

\bibitem{Muller} B.~M\"uller, {\em nucl-th/0208038}.

\bibitem{blaizot} J.P.~Blaizot, J.Y.~Ollitrault,
\Journal{\PRL}{77}{1703}{1996}.

\bibitem{DKMT} Y.~Dokshitzer, V.~Khoze, A.~Mueller, S.~Troyan,
`Basics of perturbative QCD', Editions Frontieres (1991).

\bibitem{GGX02} K.~Gallmeister, C.~Greiner, Z.~Xu,
{\em nucl-th/0202051}, proceedings of the Int. Workshop
XXX on Gross Properties of Nuclei and Nuclear Excitations
(Hirschegg 2002).

\bibitem{gavin}
S.~Gavin, R.~Vogt, \Journal{\NPB}{345}{104}{1990};
W.~Cassing, E.~Bratkovskaya, \Journal{\NPA}{623}{570}{1997}.

\bibitem{jochen}
J.~Geiss, C.~Greiner, E.~Bratkovskaya, W.~Cassing, U.~Mosel,
\Journal{\PLB}{447}{31}{1999}.

\bibitem{Fritiof} H.~Pi, {\em Comp.~Phys.~Commun.} {\bf 71} (1992), 173.

\bibitem{PDG} K.~Hagiwara et al.~[PDG], \Journal{\PRD}{66}{010001}{2002}.

\bibitem{PYTHIA} T.~Sj\"ostrand et al., {\em Comp.~Phys.~Commun.} {\bf 135},
238 (2001).

\bibitem{Albajar90} C.~Albajar et al.~[UA1],
\Journal{\NPB}{335}{261}{1990}.

\bibitem{Ex130} A.~Drees, \Journal{\NPA}{698}{331}{2002}.


\bibitem{PHENIX2} W.~Zajc et al.~[PHENIX],
\Journal{\NPA}{698}{39}{2002}.

\bibitem{Wang2} E.~Wang and X.N.~Wang,
\Journal{\PRL}{87}{142301}{2001}.

\bibitem{Bass} S.A.~Bass, \Journal{\NPA}{661}{205}{1999}.

\bibitem{referee} We thank the referee for pointing this out.

\bibitem{DokKha01} Yu.~L.~Dokshitzer, D.~E.~Kharzeev,
\Journal{\PLB}{519}{199}{2001}.

\bibitem{Munshi} M.~G.~Mustafa, D.~Pal, D.~K.~Srivastava, M.~Thoma,
\Journal{\PLB}{428}{234}{1998} 

\bibitem{PhenixEl} K.~Adcox et al.~[PHENIX],
\Journal{\PRL}{88}{192303}{2002}

\bibitem{KaiQM} K.~Gallmeister, B.~K\"ampfer, O.~P.~Pavlenko, 
{\em nucl-th/0208006}.

\end{thebibliography}
\end{document}